# CMS-XSEN:
# LHC Neutrinos at CMS.
# Experiment Feasibility Study

S. Buontempo[a], G.M. Dallavalle[b], G. De Lellis[c], D. Lazic[d], F. L. Navarria[e]

**Abstract**

We discuss a CMS eXtension for Studying Energetic Neutrinos (CMS-XSEN). Neutrinos at the LHC are abundant and have unique features: their energies reach out to the TeV range, and the contribution of the $\tau$ flavour is sizeable. The measurement of their interaction cross sections has much physics potential. The pseudorapity range $4<|\eta|<5$ is of particular interest since leptonic W decays provide an additional contribution to the neutrino flux from b and c production. A modest detector of $4.1\text{x}10^{27}$ nucleons/cm$^2$, placed in the LHC tunnel, 25 m from the interaction point, around the focusing magnet (Q1) closest to CMS, can cover that region. The hadronic calorimeter HF and the CMS forward shield will protect it from the debris of pp collisions. With a luminosity of 300/fb, foreseen for the LHC run in the years 2021-2023, the detector can observe over a thousand $\tau$ neutrino interactions, and a hundred TeV-neutrino interactions of all flavours. Several backgrounds are considered. A major source can be prompt muons from the interaction point. However, the CMS magnetic field and the structure of the Forward Shield make the estimation of their flux in the location of interest uncertain. Besides, machine induced backgrounds are expected to vary rapidly while moving along and away from the beam line. We propose to acquire experience during the 2018 LHC run by a brief test with a small Neutrino Experiment Demonstrator, based on nuclear emulsions.

(a) INFN, sezione di Napoli, Italy
(b) INFN, sezione di Bologna, Italy
(c) Università di Napoli “Federico II” and INFN, sezione di Napoli, Italy
(d) Boston University, USA
(e) Dipartimento di Fisica dell’Università and INFN, sezione di Bologna, Italy

## 1. INTRODUCTION

Neutrinos are abundant in LHC interactions. They arise promptly in W and Z leptonic decays, and from b and c decays.LHC neutrinos are unique: their energy extends to the TeV range, and the contribution of $\tau$ flavour is sizeable [1, 2].

Neutrino masses and mixing imply physics beyond the SM. At some high energy there can be non-standard neutrino interactions. TeV-neutrino interactions in the laboratory are uncharted physics (Figure 1). Interactions of muon neutrinos of cosmic origin with energies beyond 6 TeV have recently been recorded [4]. LHC can fill in the gap, and test different neutrino flavours.

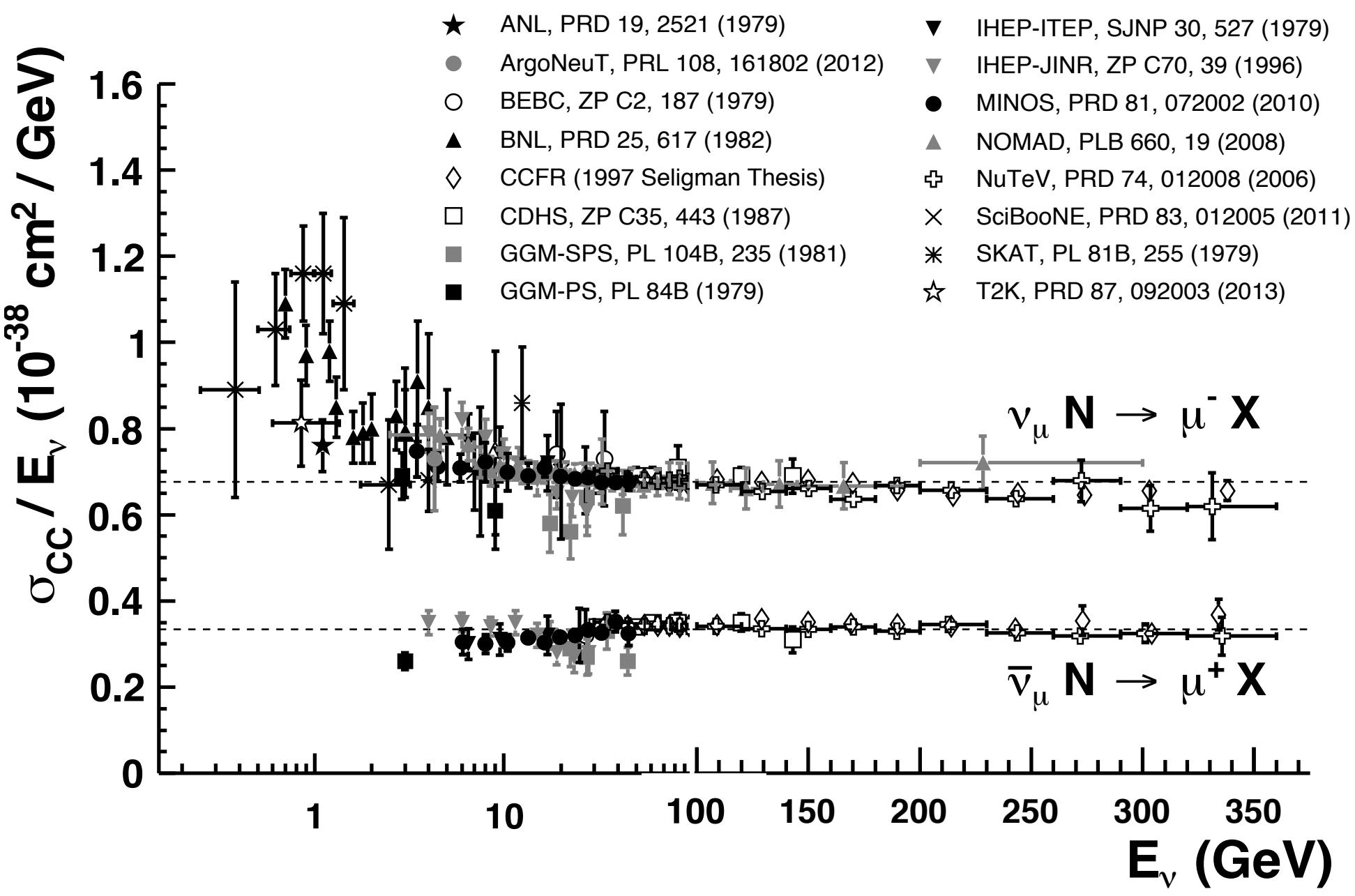

Figure 1. PDG 2016 compilation [3]

$\tau$ neutrinos are of particular interest. Only a handful of $\nu_\tau$ interactions have been directly observed so far (DONUT,OPERA[5]).

Hints at anomalies in the $\tau$ sector have accumulated:

- the precise LEP data measured an excess of the branching ratio $W \rightarrow \tau\nu_\tau$ with respect to the other leptons at the level of 2.8 sigmas [6], which persists [3];

- the Babar, Belle, LHCb experiments observe a τ excess (>3 sigmas) in semileptonic decays of B in D, D* [7].

## 2. FEATURES OF LHC NEUTRINOS

Proton-proton collision events at √s = 13 TeV were simulated using Pythia 8226 [8]. Figure 2 is a scatter plot of energy versus pseudo-rapidity η for neutrinos from leptonic W decays and from b and c production. Forward neutrinos are more energetic because of the Lorentz boost along the beam axis. Neutrinos from heavy quarks are abundant, but are very forward — a region with sizeable beam background — and the τ neutrinos account for about 5% of the neutrino flux. In the region of 4<|η|<5 a consistent contribution to the TeV neutrino tail comes from leptonic W decays. A third of those are τ neutrinos.

The W is boosted along the beam line; the charged lepton and the neutrino from the decay tend to be on the same side in pseudo-rapidity, and they are back-to-back in the plane transverse to the beam. When the neutrino lies in 4<|η|<5, the charged lepton is within the CMS acceptance (|η|<3) in about 50% of the cases. These features could open a unique possibility of pre-tagging the neutrino species.

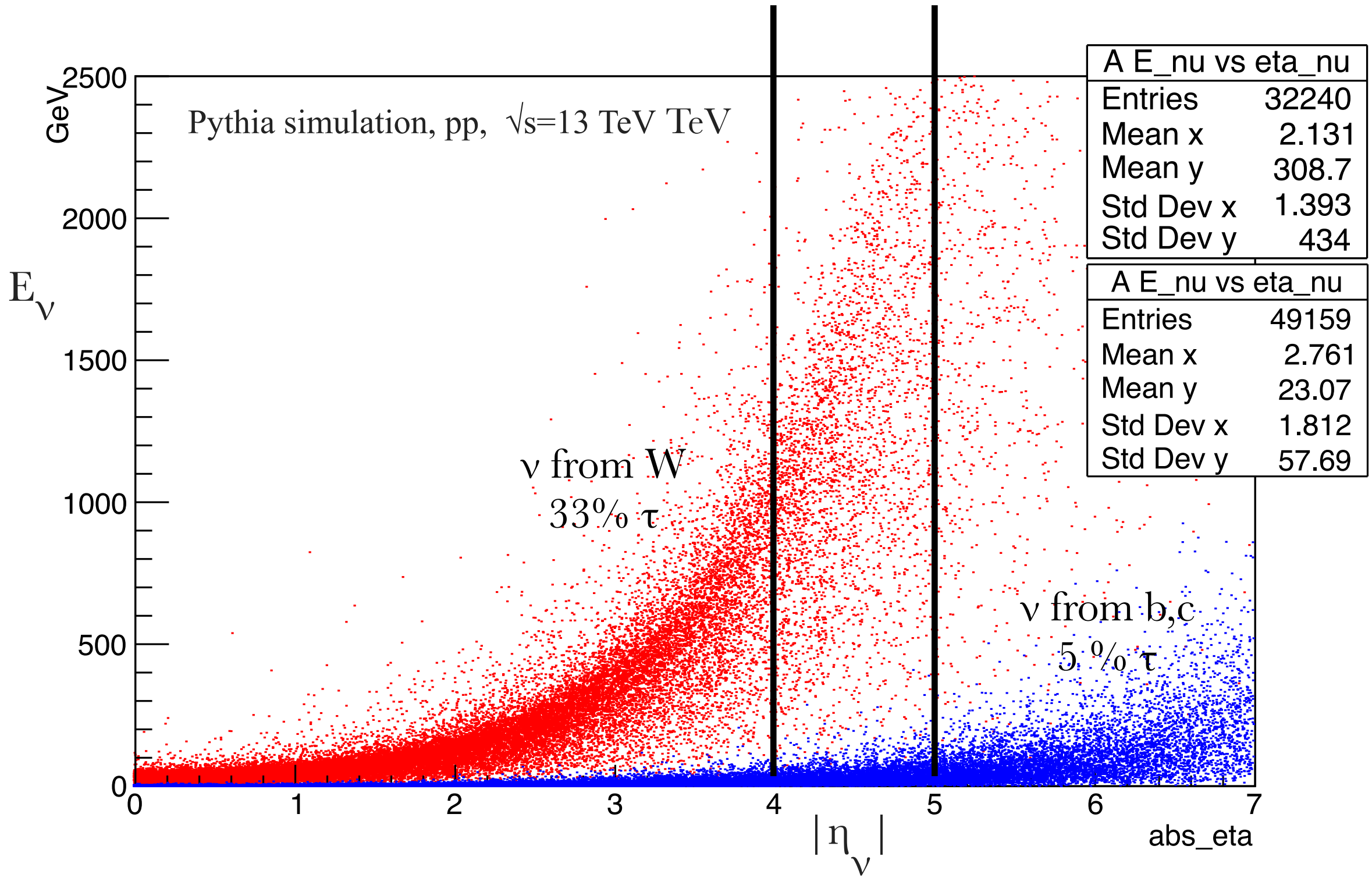

Figure 2. Pythia 13 TeV pp collisions

It is remarkable that the area surrounding the focusing magnet (Q1) closest to CMS in the LHC tunnel matches the 4<|η|<5 acceptance (Figure 3). In the following we assume a hypothetical cylindrical detector (XSEN) placed at 25 m from the Interaction Point with radius between 35 and 90 cm.

The CMS forward hadronic calorimeter (HF) — 10 nuclear interaction lengths — and the CMS forward shield — composed of steel and borated concrete, with a total of 26 nuclear interaction lengths — provide an excellent shield against all particles emerging from the interaction point (IP), but muons and neutrinos.

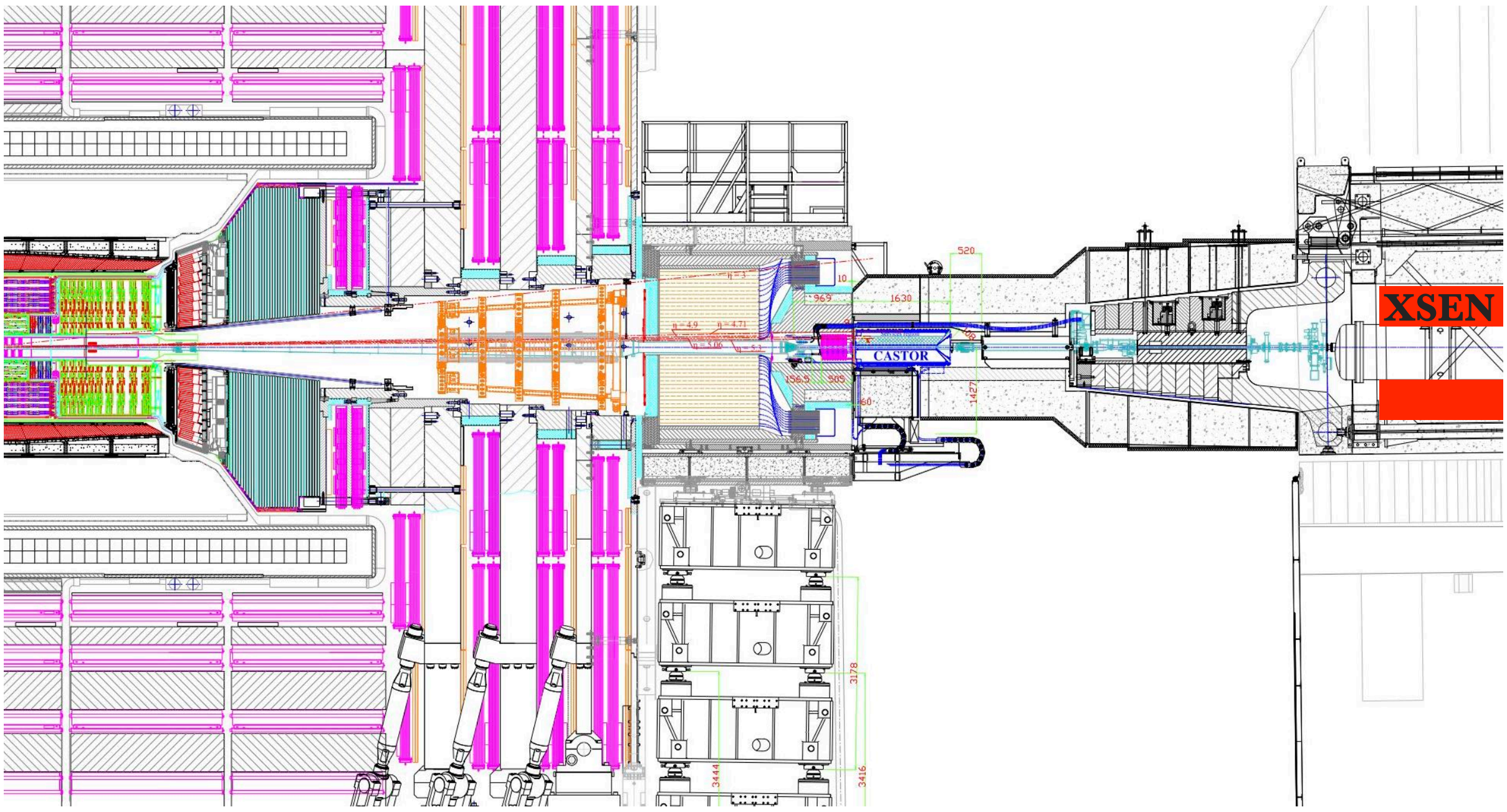

Figure 3. Hypothetical detector (XSEN) surrounding Q1

## 3. STATISTICS OF NEUTRINO INTERACTIONS

Neutrino fluences, and statistics of observable interactions in XSEN for the configuration of section 2 are summarised in Table 1. Over a thousand τ neutrino interactions, and a hundred TeV-neutrino interactions can be observed, with a luminosity of 300/fb and a modest detector of $4.1 \times 10^{27}$ nucleons/cm$^2$.

Table 1. Observable neutrino interactions in XSEN. Labels are discussed in the text. In (v) 980 are neutrinos, 490 anti-neutrinos; in (iv) 34 ν, 18 anti-ν.

| | ***$W^+ \rightarrow l^+ \nu$ (l=e,μ,τ)*** | ***$W^- \rightarrow l^-$ vbar (l=e,μ,τ)*** | **b,c production** |
|---|---|---|---|
| production Cross Section x BR (nb) | 11.8 (x3) (o) | 8.8 (x3) (o) | $48.0 \times 10^5$ (Pythia) |
| neutrinos in $4<abs(\eta)<5$ | 13% (pythia) | 2.7% (pythia) | 5.8% |
| # for 300/fb (install in LS2, until LS3) | $1.3 \times 10^9$ | $2.1 \times 10^8$ | $0.83 \times 10^{14}$ |
| # of $\nu_\tau$ | $4.5 \times 10^8$ | $7.1 \times 10^7$ | $4.5 \times 10^{12}$ (50% νbar) |
| $\langle E_\nu \rangle$ GeV (RMS) | 1200 (500) | 1000 (500) | 30 (30) |
| $\nu_\tau N$ cross section $10^{-38}$ $cm^2$ (i) | 650 | 320 | 12 (ν) 6 (νbar) |
| # of $\nu_\tau$ DIS events in tagger (ii) | 11 (iii) | 1 (iii) | 1470 (v) |
| # of ν DIS events >500GeV in tagger (ii) (all flavours) | 27 | 2 | 52 (iv) |

It is expected that the LHC will deliver such an amount of luminosity in the three year run (2021-2023) between Long Shutdown 2 (LS2) and the following LS3.

For the calculation of fluences, events have been generated with Pythia 8.2.2.6 at $\sqrt{s}$= 13 TeV. These are used to estimate the geometrical acceptance while the process cross section is rescaled to the measured values [9] ((o) in Table 1). The differential distributions are in excellent agreement with the measurements, in particular for the pseudo-rapidity distribution of leptons from W decays, which has a different shape for $W^+$ and $W^-$ . Leptons from $W^-$ are more central.

For calculating the rate of neutrino interactions, we have assumed a detector featuring $4.1 \times 10^{27}$ nucleons/cm$^2$, like 6 meters of lead, interleaved with emulsions and fine scintillators or gas ionisation chambers (tagger (ii) in Table 1). It should be noted that at high energy the active layers could be less numerous: a $\tau$ of 1 TeV has an average track length of 5 cm. A high position accuracy is needed to observe the $\tau$ decay vertex, in particular in single prong.

For the neutrino-nucleon interaction cross section, the linear dependence on energy as shown in Figure 1 is modified at high energies by the W propagator. Measurements exist by the IceCube collaboration for cosmic muon neutrinos with energies exceeding 6 TeV [4]. We have assumed an isoscalar nucleon charged current (CC) cross section of 325 and of 175 times $10^{-38}$ cm$^2$ respectively for muon neutrinos and antineutrinos of 500 GeV.

For the CC cross section of $\tau$ neutrinos on nucleons, we can only refer to theoretical expectations [10] ((i) in Table 1). The predictions are smaller than for muon neutrinos, due to the effect of the $F_4$ and $F_5$ structure functions [11].

## 4. BACKGROUNDS

Measurements by LHCb [12] and TOTEM [13] show that in $4<|\eta|<5$ there are about 4.5 charged particles on average per pp inelastic interaction at $\sqrt{s}$ = 13 TeV, mostly pions and kaons. With a pp inelastic cross section of 80 mb and a luminosity of 2 x $10^{34}$ cm$^{-2}$ s$^{-1}$, it is estimated that in $4<|\eta|<5$ there are 7x $10^9$ charged particles per second, and a total of $10^{17}$ for the 300 /fb expected in the LHC Run 3. However, the detector XSEN of Figure 3 is shielded from charged hadrons emerging from the IP by the CMS hadronic calorimeter (HF) and the CMS forward shield, providing together 36 nuclear interaction lengths, i.e. a dumping factor of $2.3 \times 10^{-16}$, corresponding to almost full absorption.

Muons from pion and kaon decays will be rare: $\pi$ and K with energies larger than a few GeV will interact before they can decay, and at lower energies the low momentum muon from the decay will not penetrate the forward shield.

The muon rate from pp collisions at $\sqrt{s}$ = 13 TeV in $4<|\eta|<5$ was measured by LHCb [14]: in Minimum Bias events they observe a rate of 25 kHz of muons with p>10 GeV at a luminosity of 4 x $10^{32}$ cm$^{-2}$ s$^{-1}$. This is in excellent agreement with Pythia calculations, assuming that the dominant source is b and c decays: the production cross section is $48 \times 10^5$ nb and 1.4% of these events have a muon in $4<|\eta|<5$, which gives an expectation of 27 kHz. The predicted muon rate at $2 \times 10^{34}$ $cm^{-2}$ $s^{-1}$ is 1.25 MHz, however, when considering the muon energy spectrum, only <10 % of those muons are expected to reach the detector XSEN, due to the CMS magnetic field and shielding material. Gas ionisation chambers as active layers could sustain the rate.

In the proposed location, XSEN will present a cross section of about $2.2 \times 10^4$ $cm^2$; about 6 muons per $cm^2$ will impinge on it in a second: at that rate emulsions can stand a few inverse femtobarns. This opens the possibility of profiting from the 2018 LHC luminosity ramp-up period and the early running until Technical Stop 1 for a better understanding of the background: an emulsion-based detector could be installed, and extracted at the first occasion.

The machine induced background varies very rapidly while moving along and away from the beam line. An occupancy of a few per $cm^2$ per second is estimated for muons in the beam halo [15] moving towards CMS. The flux of neutrons and photons is estimated to range from $10^{-3}$ to 1 $/cm^2/s$ [15]. The test detector suggested above and based on emulsions can be protected with borated polyethylene and lead.

A considerable contribution might come from hadronic secondaries emerging from the pp collisions at lower angles and interacting with the machine elements, like the collimators positioned at 18 m from the CMS IP [16]: the energy and angular distributions of those debris have large uncertainties. The proposed location should lie in a local minimum.

## 5. THE NEUTRINO EXPERIMENT DEMONSTRATOR

Instead of relying on simulations of the background and corresponding transport of the products through matter, a simple way to estimate the backgrounds is to do a preliminary measurement. We therefore propose to expose a simple stack of emulsions (NED) during the 2018 LHC luminosity ramp-up period and the early running until Technical Stop 1 , in order to study the amount of background in the preferred pseudorapidity region. The NED would be extracted at the first occasion, after a few hundreds of picobarns of LHC collisions. If the results of the first round of tests indicate that the detected backgrounds allow more precise measurements, we may foresee installation of another stack of emulsions during one of the following Technical Stops.

The NED consists of 10 layers of emulsions interleaved with 1 mm lead sheets. The stack is protected on all sides from neutrons and photons by a 30 mm thick layer of borated polyethylene and 5 mm of lead. It is placed underneath Q1, at about 25 m from the CMS IP , covering the region of 80 to 90 cm from the beam axis, i.e. at pseudorapidity about 4.

## 6. SUMMARY

The LHC is a unique environment for studying neutrino interactions in two aspects: $\tau$ neutrinos and TeV neutrinos. They are produced in b and c decays and in leptonic W decays. Over a thousand $\tau$ neutrino interactions, and a hundred TeV-neutrino interactions can be observed, with a luminosity of 300 /fb (planned for the LHC RUN 3 2021-2023) and a modest detector of $4.1 \times 10^{27}$ nucleons/cm$^2$, located in the LHC tunnel at 25 m from the CMS interaction point (CMS-XSEN). The detector surrounds the focusing magnet (Q1) closest to CMS and covers the pseudorapidity range $4<|\eta|<5$. It is shielded from charged hadrons emerging from the IP by the CMS hadronic calorimeter (HF) and by the CMS forward shield, providing together 36 nuclear interaction lengths. The rate of muons from the IP reaching that location is estimated to be tolerable, although large, but the calculation has considerable uncertainties. Machine induced backgrounds and contributions from hadronic secondaries at lower angles interacting with the machine elements vary very rapidly when moving along and away from the beam line.

We propose to install a small emulsion based detector (NED) at the beginning of the 2018 LHC run in order to perform a preliminary measurement of the background at the location of interest for XSEN.

## ACKOWLEDGEMENTS

We thank I.Ajguirey, A.Ball, A.Benvenuti, V.Cafaro, A.Dabrowski, D.Dattola, A. Di Crescenzo, F.Gasparini, V.Giordano, C.Guandalini, V.Klyukhin, M. Komatsu. S.Mallows, I.Redondo, V.Tioukov, W.Zeuner.